\documentstyle[12pt]{article}
\input epsf.sty
\begin{document}
\title{
{\bf \tt grc$\nu\nu\gamma$}: 
Event generator for the single- and double-photon emission 
associated with neutrino pair-production
}
\author{ 
Y.~Kurihara\footnote{yoshimasa.kurihara@kek.jp},
J.~Fujimoto\footnote{junpei.fujimoto@kek.jp}, 
T.~Ishikawa\footnote{tadashi.ishikawa@kek.jp},
Y.~Shimizu\footnote{yoshimitzu.shimizu@kek.jp} \\ 
{\small \it High Energy Accelerator Research Organization(KEK),}\\
{\small \it Tsukuba, Ibaraki 305-0801, Japan}\\ \\
T.~Munehisa\footnote{munehisa@hep.esb.yamanashi.ac.jp}\\
{\small \it Yamanashi University, Kofu, Yamanashi 400-8510, Japan}
}
\date{}
\maketitle
\begin{abstract}
A new event generator is proposed for two processes 
$e^+e^- \rightarrow \nu \bar{\nu} \gamma$
and $e^+e^- \rightarrow \nu \bar{\nu} \gamma \gamma$ where $\nu\bar\nu$
includes all the neutrino species.
The exact matrix elements of single- and double-photon
emission, generated by the {\tt GRACE} system, are convoluted with the
QED parton shower({\tt QEDPS}) to deal with the initial state 
radiations(ISR). It is pointed out that a careful treatment is required
to avoid the double counting of the radiative photons between 
the matrix elements and the ISR part. A detailed comparison of
{\tt grc$\nu\nu\gamma$} with the ${\cal O}(\alpha)$ calculations and 
other similar Monte Carlo generators is discussed on the total cross 
section and on various distributions. 
It is also examined how the possible effects of the anomalous
triple-gauge-boson couplings can be observed.
\end{abstract}
\section{Introduction}
The processes of single- and double-photon emission associated 
with a large missing
energy are very important for the precision test of the standard model
and the search for any new (invisible) particles beyond 
the standard model.
In fact the number of the light-neutrino generations has been measured 
using this process on the $Z$-pole at the LEP experiments\cite{LEP1}.
The efforts to find new invisible particles, such as a neutral 
(lightest) SUSY particle or a sequential heavy neutrino, are now
continued by the LEP-2 experiments at higher energies\cite{LEP2}. 
In this case, however, the 
neutrino pair-productions with hard-photon emission are the main
background for the investigation.
With the increase of the accumulated luminosity, the cross 
sections of these processes must require more precision
with an uncertainty of $\cal{O}$(1\%). In addition since 
$e^+e^- \rightarrow \nu_e \bar{\nu}_e \gamma (\gamma)$ includes the
triple(quartic) gauge couplings, there is a possibility to find some 
new effects beyond the standard model on these rather poor-tested 
couplings.

There are some methods to estimate the cross sections including 
the higher 
order QED corrections. The simplest one would be to combine the matrix 
elements of $e^+e^- \rightarrow \nu \bar{\nu}$ with 
some tool for the initial-state radiations(ISR).
The hard photon(s) may be supplied, for instance, by
a parton shower, which
can treat the photons with a finite transverse momentum($p_T$). 
When this method is applied to the center-of-mass system(CMS) 
energies above the $Z$-boson mass, however, 
the leading logarithmic (LL) approximation is not sufficient.
This is because the colliding energy after the photon emission 
can match the $Z$-pole('Radiative Return'), and
the probability to emit high-energy photon(s)
is largely enhanced. When this happens the LL approximation
is not successful in describing the spectrum around the $Z$ pole.
To improve the precision of the calculations the non-LL terms for ISR
must be taken into account as well as the LL terms. 
However, there exist no complete calculations up to the non-LL order.

Another possible way is to make use of the matrix elements 
of $e^+e^- \rightarrow \nu \bar{\nu} \gamma (\gamma)$\footnote
{
The complete matrix elements of the $e^+e^- \rightarrow \nu_\mu 
\bar{\nu_\mu} \gamma$
process and their compact approximated form were first given 
in Ref.\cite{berends}.
}
together with the ISR tool.
Diagrams of the single(double)-photon emission are shown in
Fig.1 (Fig.2), respectively. 
The visible and invisible hard-photon(s) are fed by 
the matrix elements as well as by the ISR tools. Here
it is crucial to avoid the double counting in the radiations 
coming from these two different sources.
The conventional QED structure function is not suitable for this 
purpose because it is obtained after the integration over all
allowed phase-space of the emitted photons except for their energy.
Thus the double counting necessarily happens and cannot be excluded. 

There exists no full electroweak one-loop calculation for 
$e^+e^- \rightarrow \nu \bar{\nu} \gamma$. 
Only the limited corrections, one-loop QED corrections and self-energy 
corrections to the $Z$-exchange diagrams\cite{berends,1-loop,IN}
are available. Though there is no one-loop calculation at all for the 
$\nu_e \bar{\nu}_e \gamma$ process, 
the ISR tools might cover the most important corrections and allows us
to give a realistic event generator which can meet the
required theoretical uncertainty of $\cal{O}$(1\%).

The {\tt KORALZ} Monte Carlo program\cite{koralz} exploits 
the exact matrix
elements of the single-photon processes with $\nu_\mu$ and $\nu_\tau$ 
production(an approximation is made for $\nu_e$) together
with the exclusive exponentiation of YFS\cite{yfs} as the ISR tool.
Another event generator is {\tt NUNUGVP}\cite{nunugpv}, which
uses the exact matrix elements of the single- and 
multi(up to three)-photon emission(for all kinds of neutrinos) 
generated by the ALPHA algorithm\cite{alpha}
and the $p_T$-depending structure function\cite{SFt} for ISR. 
Both programs include anomalous triple-gauge-boson  coupling(TGC) in 
the package.

In this report a new event generator {\bf \tt grc$\nu\nu\gamma$}
is proposed. It adopts the exact matrix elements for
$e^+e^- \rightarrow \nu \bar{\nu} \gamma (\gamma)$ 
generated by means of the {\tt GRACE} system\cite{GRACE} combined with 
{\tt QEDPS}\cite{QEDPS} for ISR.
The advantages of these packages are:
\begin{itemize}
\item The exact matrix elements up to the double-photon emission, 
including the $\nu_e$ process, are used. 
Double-photon emission is practically sufficient for experimental 
analysis.
\item {\tt QEDPS} keeps the complete kinematics for the emitted 
photons and virtual electrons before collisions. It allows a more 
flexible treatment of the ISR effects in avoiding the double-counting.
\end{itemize}
In section 2, the calculation method, particularly how to connect
{\tt QEDPS} to the radiative processes without double counting, 
is explained. The numerical results of single- and double-photon 
emission are presented in section 3. A detailed comparison among 
{\bf \tt grc$\nu\nu\gamma$},
the exact ${\cal O}(\alpha)$ calculations, {\tt KORALZ} and 
{\tt NUNUGPV} is also 
given in this section. The effects of the anomalous TGC
in the single-photon events are discussed in section 4.
The conclusions are summarized in section 5.
Appendix describes how to obtain and use the 
{\bf \tt grc$\nu\nu\gamma$} program.
\section{Calculation Method}
The exact matrix-elements of all the processes are produced by
an automatic calculation system {\tt GRACE}. It
generates the FORTRAN routines needed to calculate
the amplitudes numerically based on the helicity-amplitude formalism.

For ISR the parton shower algorithm for 
QED, {\tt QEDPS}, is used. In this algorithm the Altarelli-Parisi 
equation is solved in the LL approximation\cite{AP} using the 
Monte Carlo method. The details of this
method can be found in Ref.\cite{QEDPS}. Here we recall that the 
algorithm can maintain the exact kinematics during the evolution
of an electron. Let us consider the branching process
\begin{eqnarray}
e^-(x,K^2) \rightarrow e^-(xy,K_1^2)+ \gamma(x(1-y), Q_0^2), \nonumber
\end{eqnarray}
where the parent electron, whose virtuality is $K^2$, has the momentum 
$p=(E,{\bf{0}}_T,p_z)$,
$E=\sqrt{p_z^2-K^2}$ and $p_z=xp^*$ with $p^*$ being the momentum of the
system in the infinite-momentum-frame.
A cutoff mass $Q_0$ is introduced for the photon.
Then the momenta of the daughter particles, $p_1$ for $e^-$ and
$p_2$ for $\gamma$, can be expressed as
\begin{eqnarray}
p_1&=&(E_1,{\bf{k}}_T,yp_z), \nonumber \\
p_2&=&(E_2,-{\bf{k}}_T,(1-y)p_z), \nonumber 
\end{eqnarray}
where
\begin{eqnarray}
E_1&=&\sqrt{y^2p_z^2+{\bf k}_T^2-K_1^2}, \nonumber \\
E_2&=&\sqrt{(1-y)^2p_z^2+{\bf k}_T^2+Q_0^2}. \nonumber
\end{eqnarray}
Assuming $p^* \rightarrow \infty$, we have
\begin{eqnarray}
 -K^2 = -K_1^2/y + Q_0^2/(1-y) + \mbox{{\bf k}}^2_T/(y(1-y)), \nonumber
\end{eqnarray}
which determines ${\bf k}_T^2$ from $y$, $K^2$ and $K_1^2$.
Hence one can obtain the distribution of the transverse momentum of
the emitted photons, ${\bf k}_T^2$. Moreover this enables us
to find the virtuality of the 
electron, $-K^2$, as well as its energy, just before the collision.

The phase-space integration of the matrix element squared is 
carried out numerically by {\tt BASES}\cite{BASES} using an adaptive 
Monte Carlo method. The integration is done in the 
five(eight)-dimensional kinematical phase-space of the three(four)-body 
final state. The evolution of the initial-state electrons
by {\tt QEDPS} is treated independently of {\tt BASES}.

Since the {\tt QEDPS} can provide a complete kinematical information 
about the emitted photons and the virtual electrons, it is easy 
to distinguish
the photons from the matrix-elements and those from the parton shower. 
Let us divide the full one-photon phase space $\Omega_\gamma$ 
into two regions: One is for the $visible$ region, $\Omega_{\tt v}$, 
and the other is for the $invisible$ region,
$\Omega_{\tt i}$. These are defined as
\begin{eqnarray}
\Omega_{\tt v}&=&\{(E_\gamma,\theta_\gamma)|E_\gamma>E_0,
\theta_0 < \theta_\gamma < 180-\theta_0\}, \label{visible} \\
\Omega_{\tt i}&=&\Omega_\gamma-\Omega_{\tt v}, \nonumber
\end{eqnarray}
where $E_\gamma$ is the energy of the photon and $\theta_\gamma$ is its
polar angle in degree from the beam axis. 
The minimum energy and angle, $E_0$ and
$\theta_0$, must be given by the experimental condition.
First the photons from {\tt QEDPS}, $\gamma_{\tt ps}$'s, are restricted 
to go inside of $\Omega_{\tt i}$ and the photon from the 
matrix-elements, $\gamma_{\tt ME}$, is emitted into $\Omega_{\tt v}$.
Since there is no overlap between these two regions no double-counting
occurs. In addition the ordering of the electron virtuality
is also required. During the evolution of an electron the virtuality
is monotonically increasing, which is realized naturally in the 
{\tt QEDPS}
algorithm. A further condition must be imposed on the virtuality of 
the electron in the matrix-elements after emitting the photon: 
It  should be greater than the virtuality of the electron in 
the last stage of {\tt QEDPS}.

The configuration mentioned here cannot cover all the configurations
for the photon emission. When $n$ photons are emitted from one electron 
line, the first ($n-1$) photons are fed by {\tt QEDPS}, including 
the statistical factor of $1/(n-1)!$, and only the last photon is 
supplied by the matrix element. This configuration represents 
the case where the first ($n-1$) $\gamma_{\tt ps}$'s are invisible
but only the last one, $\gamma_{\tt ME}$, is visible. Another case 
also exists. One of the first ($n-1$) $\gamma_{\tt ps}$'s goes
into $\Omega_{\tt v}$ while the other ($n-2$) $\gamma_{\tt ps}$'s 
and $\gamma_{\tt ME}$ escape into $\Omega_{\tt i}$.
Though $\gamma_{\tt ME}$ is invisible in this case, the results are free
from any infrared divergence thanks to the virtuality cut $Q_0$ 
introduced above. The minimum energy of $\gamma_{\tt ME}$ is determined 
from the minimum virtuality imposed by the {\tt QEDPS} evolution.
If there is no 'Radiative Return' the contribution of this second 
configuration is negligible small. However, for the process
$e^+e^- \rightarrow \nu \bar{\nu} \gamma (\gamma)$ above the $Z^0$
threshold its contribution to the total cross section becomes sizeable, 
typically about a few \%.

Besides the above-mentioned pure QED corrections 
another class of electro-weak higher order corrections should be 
included in the calculations.
The {\bf \tt grc$\nu\nu\gamma$} package has two schemes to do this:
The running coupling constant and the $G_\mu$ scheme\cite{gmu}.
In the former the coupling constant of 
the $fermion$-$fermion$-$Z$ vertex, $g_{ffZ}$,
is determined by evolving it from zero momentum transfer 
to the mass squared of the $\nu\bar\nu$ system, $q_Z^2$, which differs 
from one event to another. It varies according to the
renormalization group equation (RGE) as
\begin{eqnarray}
g_{ffZ}(q_Z^2)=g_{ffZ}(0)\left(1-\frac{\alpha}{3\pi}
\sum_i C_i e_i^2 \log{q_Z^2\over m_i^2} \right)^{-1}, \nonumber
\end{eqnarray}
where $\alpha=1/137.036$ is the QED coupling at the zero momentum
transfer, $C_i$ the color factor, $e_i$ the electric charge, $m_i$ the
mass of the $i$'s fermion, and $i$ runs over all massive fermions.
In this scheme $g_{ffZ}(q_Z^2)$ should be
fixed event-by-event, but is common for all $Z$-exchange 
diagrams in the same event.

The latter scheme is such that the weak couplings are determined 
through the weak-mixing angle, $\sin{\theta_W}$, which is given by
\begin{eqnarray}
{\rm sin}^2 \theta_W=\frac{\pi \alpha(q^2)}{\sqrt{2} G_\mu M_W^2}
\frac{1}{1-\Delta r}, \nonumber
\end{eqnarray}
where $M_W$ being the $W$-boson mass and $G_\mu$ the muon decay 
constant. In this scheme the input is the precisely measured value 
of $G_\mu$ at zero momentum transfer and 
$\alpha(q^2)$ is the QED coupling evolved to a typical
energy scale of the processes which should be chosen by the users.
In the present version of the program the loop corrections expressed by 
$\Delta r$ in the formula is ignored($\Delta r=0$). 
Then the difference between the running $g_{ffZ}$ scheme
and the $G_\mu$ scheme is to use the fixed energy scale(for $G_\mu$) 
or the event-dependent energy scale(for running $g_{ffZ}$).

The package {\bf \tt grc$\nu\nu\gamma$} has a switch to choose either 
of the above corrections or no higher-order corrections. 
When users select the last option,
the weak couplings are simply fixed by $M_W$ and $M_Z$
through the on-shell relation,
${\rm sin}^2 \theta_W =1-\frac{M_W^2}{M_Z^2}$, 
where $M_Z$ is the mass of the $Z$-boson.
\section{Numerical Results}
\subsection{$1\gamma$ test}
The numerical results for the single-photon emission are given in 
this section. All the three generators have been run and compared.
The parameters used in the calculations are summarized in Table 1.
\begin{table}[htbp]
\begin{center}
\begin{tabular}{|c|c||c|c|}
\hline
$M_Z$ & 91.187~GeV& $\Gamma_Z$ & 2.49 GeV \\ \hline 
$M_W$ & 80.22~~GeV& $m_e$ & $0.511 \times 10^{-3}$ GeV \\ \hline
$\alpha(0)$ & 1/137.04 & $\alpha(M_W^2)$ & 1/128.07 \\ \hline
\end{tabular}
\end{center}
\caption{\footnotesize Parameters used in the calculations.}
\end{table}

The weak coupling constant is determined scheme by scheme according to 
the selection of the higher-order corrections as explained in the end of
the previous section. The $visible$ photon is defined by 
Eq.(\ref{visible}) with $E_0=1$ GeV and $\theta_0=10^\circ$.

First the results for $\nu_\mu$ are discussed. In this case four program
packages are available: the analytic $\cal{O}$($\alpha$) 
calculations,\footnote
{
The formulae given in Ref.\cite{IN} are used in this paper.
}
{\tt NUNUGPV}, the present one and {\tt KORALZ}. The last one also
uses the complete matrix-elements for this process
including the YFS exponentiation.
The cross sections from four independent generators are visualized in 
Fig.3, and the numbers at four energy points are shown in Table 2.
In these calculations $\alpha=1/137.04$ and 
${\rm sin}^2 \theta_W =1-M_W^2/M_Z^2$ 
are commonly used for all four packages.
\begin{table}[htbp]
\begin{center}
\begin{tabular}{|c|c|c|c|c|}
\hline
&160 GeV&170 GeV&180 GeV&190 GeV \\ \hline
{\bf \tt grc$\nu\nu\gamma$}& 2.470 & 2.033 & 1.709 & 1.460\\ \hline
$\cal{O}$($\alpha$) corr.  & 2.436 & 2.006 & 1.689 & 1.449\\ \hline
{\tt KORALZ}               & 2.437 & 2.009 & 1.697 & 1.459\\ \hline
{\tt NUNUGPV}              & 2.410 & 1.987 & 1.668 & 1.430\\ \hline
\end{tabular}
\end{center}
\caption{\footnotesize
Total cross sections of single-photon emission for the
$\nu_\mu$ process in $pb$. The on-shell scheme without any higher-order 
electro-weak corrections is used for all packages. 
}
\end{table}
The statistical errors to the numbers in the table are around 0.2\% 
due to the numerical integration. 
The results from {\tt KORALZ} are very close to those from 
the $\cal{O}$($\alpha$) calculation. Since
the former includes the soft-photon exponentiation 
but the latter does not, one sees that the resultant difference in
the total cross sections is less than 1\%.
In the {\tt grc$\nu\nu\gamma$} package, however, {\tt QEDPS} allows 
also the hard-photon exponentiation in the LL approximation. 
From the table this brings an effect of around 1.4\%.
By comparing three Monte Carlo program packages one can conclude that 
the theoretical uncertainty from the ISR corrections is about 1\%.
The results of {\tt NUNUGPV} are lower than 
{\bf \tt grc$\nu\nu\gamma$} by around 2\%. 

The differential cross sections with respect to the CMS energy of 
the pair $\nu \bar{\nu}$, 
its longitudinal component, energy and transverse energy of 
the hard photon
are compared in Fig.4. The shape of the distributions from these 
three packages ({\bf \tt grc$\nu\nu\gamma$},
$\cal{O}$($\alpha$), and {\tt KORALZ}) is in good agreement. 
A more
detailed comparison between {\bf \tt grc$\nu\nu\gamma$} and {\tt KORALZ}
on the same distributions is given in Fig.5, which shows the ratio of 
the results. Besides the overall factors these 
two are consistent.

The effects of the higher-order corrections due to the running $ffZ$
coupling are shown in Fig.6 and Table 3 at four energy-points.
\begin{table}[htbp]
\begin{center}
\begin{tabular}{|l|c|c|c|c|}
\hline
&160 GeV&170 GeV&180 GeV&190 GeV \\ \hline
{\bf \tt grc$\nu\nu\gamma$} ($\nu_\mu$ only)& 2.470 & 2.033 & 1.709 & 1.469\\ \hline
$\uparrow$ + running $g_{ffZ}$& 2.846 & 2.334 & 1.968 & 1.678\\ \hline
$\uparrow$ for all neutrino species & 11.16 & 9.70 & 8.62 & 7.85\\ \hline
\end{tabular}
\end{center}
\caption{\footnotesize
Total cross sections of single-photon emission for the
$\nu_\mu$ and all neutrino processes 
using {\bf \tt grc$\nu\nu\gamma$} in $pb$.  
}
\end{table}
The higher order effects from the running $g_{ffZ}$ contribute to 
the total cross section by around 13\%. After summing up all 
the neutrino species the total cross sections are also summarized 
in the figure and the table. The contribution from the $W$-exchange 
diagrams(last three diagrams in Fig.1) amounts about 25\%.

The scheme dependence of the higher-order corrections is found
by {\tt grc$\nu\nu\gamma$}
to be around 1\% as seen from Fig.7 and Table 4.
Here $q^2=M_W^2$ is chosen as the energy scale
for the $G_\mu$ scheme.
\begin{table}[htbp]
\begin{center}
\begin{tabular}{|l|c|c|c|c|}
\hline
&160 GeV&170 GeV&180 GeV&190 GeV \\ \hline
{\bf \tt grc$\nu\nu\gamma$} + running $g_{ffZ}$
& 11.16 & 9.70 & 8.62 & 7.85\\ \hline
{\bf \tt grc$\nu\nu\gamma$} + $G_{\mu}$ scheme
& 10.98 & 9.56 & 8.53 & 7.81 \\ \hline
{\tt NUNUGPV}
& 10.76 & 9.41 & 8.37 & 7.69 \\ \hline
\end{tabular}
\end{center}
\caption{\footnotesize
Total cross sections of the single-photon emission for the
all neutrino processes using {\bf \tt grc$\nu\nu\gamma$} 
and {\tt NUNUGPV} in $pb$. The latter package implements 
the $G_\mu$ scheme for the higher-order corrections. 
}
\end{table}
The discrepancy between two schemes can be understood as follows.
Two different energy scales are involved in this process:
The colliding energy of the initial $e^+e^-$ and the $Z$-boson 
mass.\footnote{There is another energy scale $q^2=0$ for the real 
photon emission.
This can be easily separated by using a fixed value of 
$\alpha=1/137.04$ at the $e$-$e$-$\gamma$ vertex in the matrix elements 
and the ISR part.
}
The contributions to the cross section from these two energy regions 
are compatible in magnitude as seen from Fig.4. An expected
difference in the total cross section could be estimated as
$1-\alpha^2(M_Z^2)/\alpha^2(E_{\rm LEP2}^2) \approx 2\%$ 
between these two scales. 
Hence the $G_\mu$ scheme which depends on a $single$ fixed energy 
scale is not suitable for those processes that involve different 
energy scales.

The differential distributions in $G_\mu$ scheme obtained by 
{\bf \tt grc$\nu\nu\gamma$} are in good agreement with those 
by {\tt NUNUGPV} as shown in Fig.8.

\subsection{$2\gamma$ test}
For the double-photon emission a similar comparison 
is also made between {\bf \tt grc$\nu\nu\gamma$} and
{\tt NUNUGPV}. Both packages equip the exact matrix-elements for
the double-photon emission associated with all neutrinos including
$\nu_e\bar\nu_e$(Fig.2) and the $G_\mu$ scheme for the higher order 
corrections but different ways for the LL exponentiation for 
ISR, {\tt QEDPS} or $p_T$-depending structure function. Hence
a difference occurs again in the treatment of the second configuration 
of the photon emission discussed in section 2. 
In applying {\bf \tt grc$\nu\nu\gamma$} to the double-photon emission, 
one has to take into account the possibility
that one of the two photons supplied by the matrix-elements can be 
invisible(in $\Omega_{\tt i}$) and one 
photon from {\tt QEDPS} becomes visible.
The configuration in which both photons from the matrix-elements
disappear into $\Omega_{\tt i}$ is ignored safely.
The total cross sections of the double-photon emission are summarized 
in Fig.9 and Table 5, where a slight difference is seen between
two packages. 
\begin{table}[htbp]
\begin{center}
\begin{tabular}{|l|c|c|c|c|}
\hline
&160 GeV&170 GeV&180 GeV&190 GeV \\ \hline
{\bf \tt grc$\nu\nu\gamma$} & 0.687 & 0.599 & 0.537 & 0.481\\ \hline
{\tt NUNUGPV}               & 0.669 & 0.578 & 0.507 & 0.472 \\ \hline
\end{tabular}
\end{center}
\caption{\footnotesize
Total cross sections of the double-photon emission for 
all neutrino processes using {\bf \tt grc$\nu\nu\gamma$} 
and {\tt NUNUGPV} in $pb$. The latter package implements 
the $G_\mu$ scheme for the higher-order corrections. 
}
\end{table}
The differential distributions are in good agreement with 
each other as shown in Fig.10.

\section{Anomalous $WW\gamma$ Coupling}
In the {\bf \tt grc$\nu\nu\gamma$} package
the anomalous coupling of the $W$-$W$-$\gamma$ vertex is prepared.
The program includes only those terms which conserve $C$ and $P$
invariance, derived from the following effective
Lagrangian\cite{hagiwara}:
\begin{eqnarray*}
L_{eff} =-ie[(1+\Delta g_{1\gamma})(W^{\dag}_{\mu \nu} W^{\mu} -
                  W^{\dag \mu} W_{\mu \nu}) A^\nu
+(1+\Delta \kappa_\gamma) W^{\dag}_{\mu}W_{\nu}A^{\mu \nu} \\
+\frac{\lambda_\gamma}{M_W^2}W^{\dag}_{\lambda \mu}
           W^{\mu}_{\nu}A^{\lambda\nu}], 
\end{eqnarray*}
where $W_{\mu\nu}=\partial_{\mu}W_{\nu}-\partial_{\nu}W_{\mu}$, 
$A_{\mu\nu}=\partial_{\mu}A_{\nu}-\partial_{\nu}A_{\mu}$.
Here $\Delta g_{1\gamma}$,
$\Delta \kappa_\gamma$ and $\lambda_\gamma$ stand for the anomalous 
coupling parameters which vanish in the standard model. 

To find the region sensitive to the anomalous TGC
in the photon phase-space, the energy and angular distributions 
are examined at $\Delta g_{1\gamma}=0$,
$\Delta \kappa_\gamma=-10$ and $\lambda_\gamma=0$. This is shown 
in Fig.11. The sensitive region, as expected from a glance at
the diagrams, locates in the high energy side and large angle region, 
except for around the 'Radiative Return', where the annihilation
dominates the process. Then the experimental cuts of 
20 GeV$<E_\gamma<$55 GeV or 65 GeV$<E_\gamma$ 
and $45^{\small o}<\theta_\gamma<135^{\small o}$ are effective to 
enhance the signal from the anomalous TGC.
The total cross section as a function of $\Delta \kappa_\gamma$ and 
$\lambda_\gamma$ with the experimental cuts at the CMS energy at 
170 GeV is shown in Fig.12. Once 200$pb^{-1}$ of the accumulated 
luminosity and 100\% of the acceptance
are assumed, $128\pm11$ events will be observed. The error given here 
is only a statistical one.
The sensitivity of $-9$ to $+6$ for $\lambda_\gamma$ and 
$-3$ to $+10$ for $\Delta \kappa_\gamma$ is expected at the
three-sigma limit if one supposes only the statistical error of 
the measurements.
Since the higher order corrections also change the total cross sections
under the above cuts, as pointed out in Ref.\cite{koralz}, a careful 
investigation is needed on the systematics from those corrections.
\section{Conclusions}
An event generator {\bf \tt grc$\nu\nu\gamma$}
for the processes $e^+e^- \rightarrow \nu \bar{\nu} \gamma$
and $e^+e^- \rightarrow \nu \bar{\nu} \gamma \gamma$
has been proposed. The exact matrix elements for single- and 
double-photon emission are prepared by {\tt GRACE}.
{\tt QEDPS} is used as a tool for the ISR corrections.
For the $\nu_\mu$  case
the total cross sections and the hard-photon distributions of
{\bf \tt grc$\nu\nu\gamma$} are compared with those from 
the $\cal{O}$($\alpha$) calculation, {\tt KORALZ} and {\tt NUNUGPV}.
It is found that the theoretical uncertainty for the ISR corrections
is under control at the 1\% level. The systematics of the $G_\mu$ 
scheme from the double energy scales involved in the reaction
is estimated to be around 1\%.
The energy spectrum of the hard-photons is in a reasonable agreement 
with independent calculations from
{\tt KORALZ} and {\tt NUNUGPV} up to the double-photon emission.

Concerning $\nu_e$ a similar comparison with {\tt NUNUGPV} has been
done, though in lacking ${\cal O}(\alpha)$ calculations.
With this process there is some opportunity to measure
the anomalous TGC at the $W$-$W$-$\gamma$ vertex which appears in the
single energetic events at the LEP-2 when the luminosity is enough
accumulated.
Since only the experimental observable is the total cross section,
a careful investigation of the systematic errors must be done.

\vskip 1cm  
The authors would like to thank  
G.~Montagna, O.~Nicrosini, and F.~Piccinini
for the various discussions on the comparison between
{\bf \tt grc$\nu\nu\gamma$} and {\tt NUNUGVP}, and S.~Jadach 
for {\tt KORALZ}.  
We also wish to thank D.~Perret-Gallix at LAPP, and 
K.~Tobimatsu and K.~Kato at Kogakuin-University for 
their fruitful discussions and suggestions. 

This work was supported in part by the Ministry of Education,
Science and Culture, Japan under the Grant-in-Aid for Scientific Research 
No.11440083.



\begin{figure}[htb]
\centerline{
\epsfysize=15cm
\epsfbox{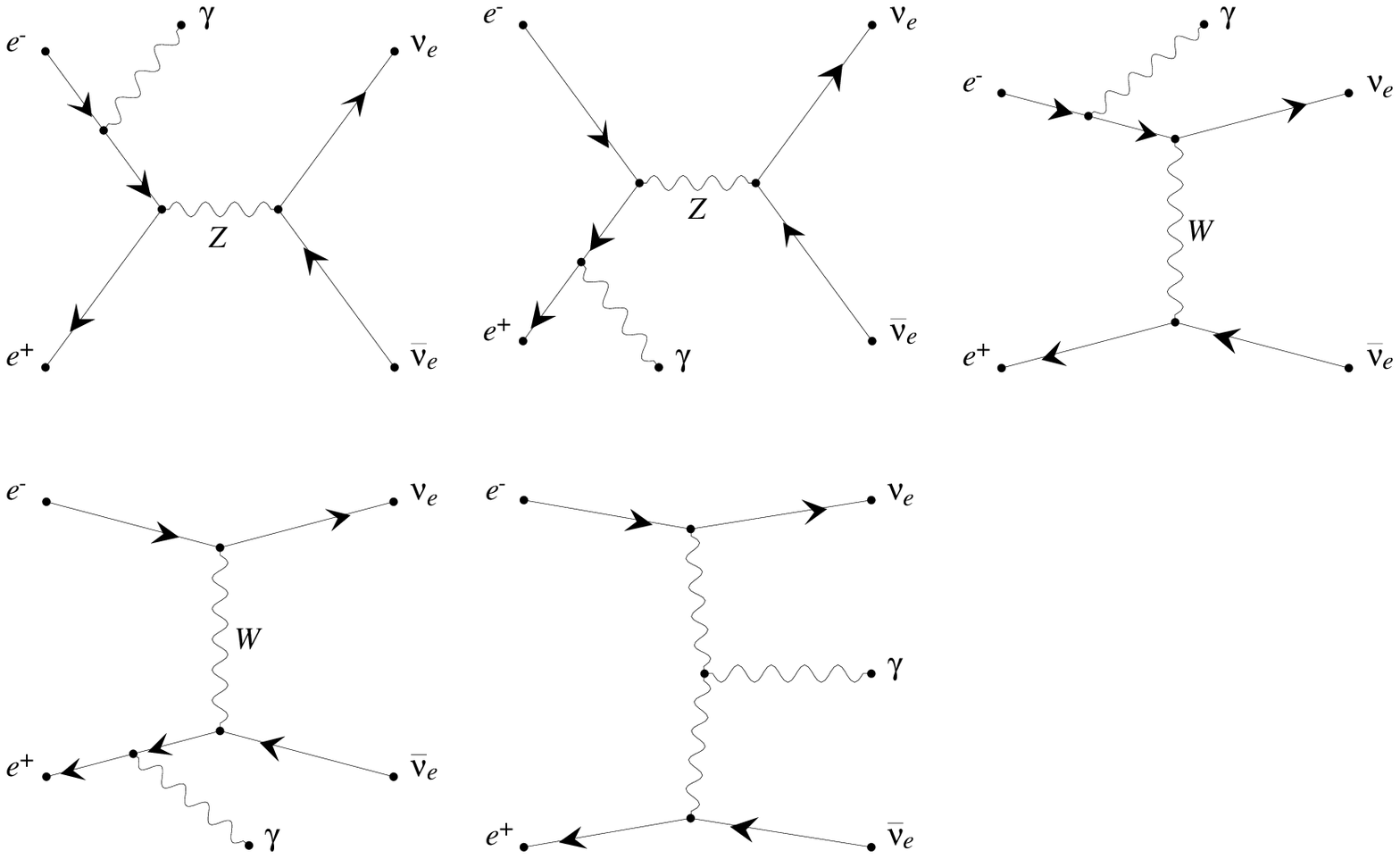}
}
\caption[FIG:diag1a]{\label{FIG:diag1a}\footnotesize
Feynman diagrams of the process
$e^+e^- \rightarrow \nu_e \bar{\nu}_e \gamma$.
Only the first two diagrams appear in $\nu_{\mu}$ 
and $\nu_{\tau}$ neutrino production. 
}
\end{figure}
\begin{figure}[htb]
\centerline{
\epsfysize=15cm
\epsfbox{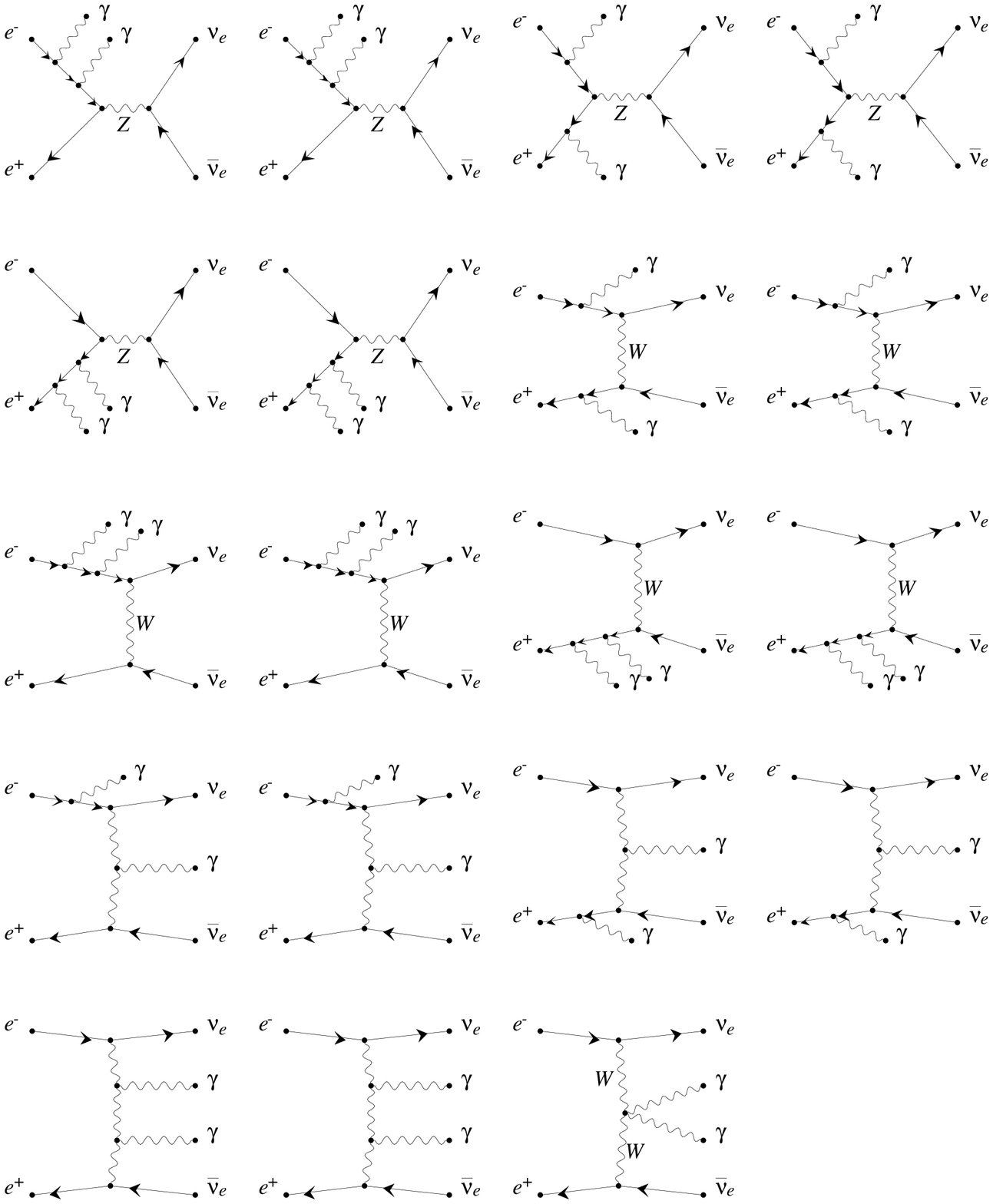}
}
\caption[FIG:diag2a]{\label{FIG:diag2a}\footnotesize
Feynman diagrams of the process
$e^+e^- \rightarrow \nu_e \bar{\nu}_e \gamma \gamma$.
Only the first six diagrams appear in $\nu_{\mu}$ 
and $\nu_{\tau}$ neutrino production. 
}
\end{figure}
\begin{figure}[htb]
\centerline{
\epsfysize=10cm
\epsfbox{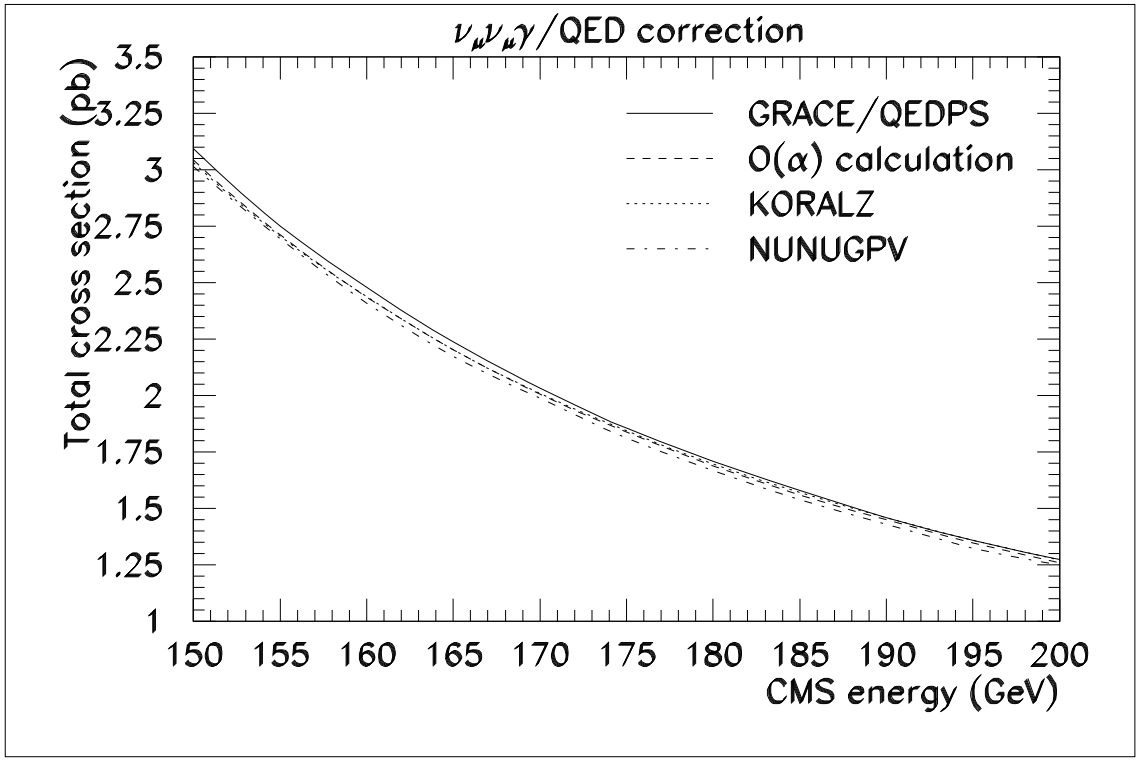}
}
\caption[FIG:roots1]{\label{FIG:roots1}\footnotesize
Total cross sections of the single-photon emission for the
$\nu_{\mu}$ process in $pb$. The on-shell scheme without any higher-order 
electro-weak correction is used for all the program packages. 
}
\end{figure}
\begin{figure}[htb]
\centerline{
\epsfysize=15cm
\epsfbox{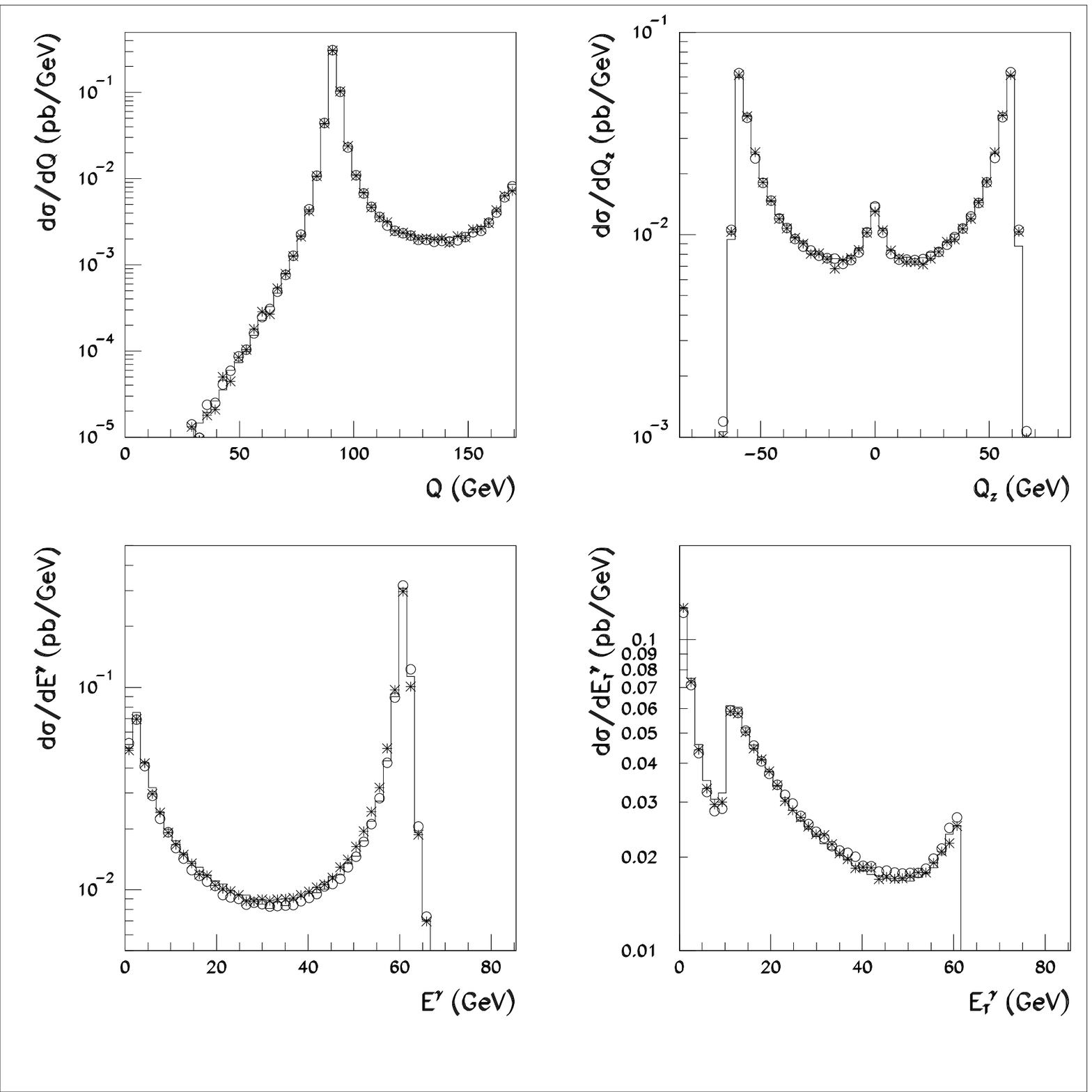}
}
\caption[FIG:spect1]{\label{FIG:spect1}\footnotesize
Differential cross sections with respect to  the CMS energy of $\nu \bar{\nu}$ system,
its longitudinal component, energy and transverse energy of the hard photon
obtained by {\bf \tt grc$\nu\nu\gamma$} (solid histograms), 
$\cal{O}$($\alpha$) calculations (stars), and {\tt KORALZ} (circles) at the CMS
energy of 170 GeV.
}
\end{figure}
\begin{figure}[htb]
\centerline{
\epsfysize=15cm
\epsfbox{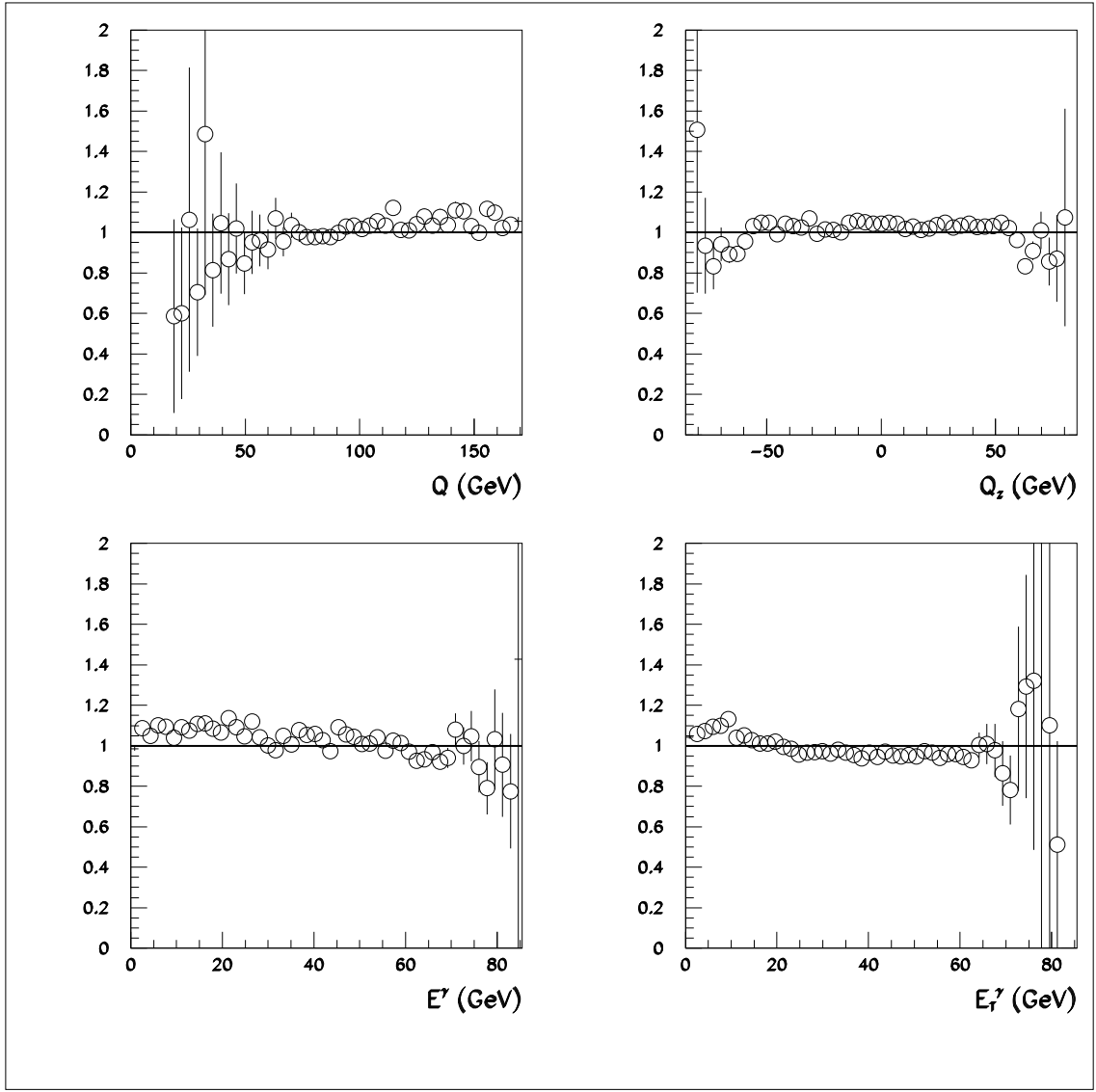}
}
\caption[FIG:spect2]{\label{FIG:spect2}\footnotesize
Ratio between {\bf \tt grc$\nu\nu\gamma$} and {\tt KORALZ} for the
same distributions as in Fig.4. The error bars show the statistical
errors of the numerical integration.
}
\end{figure}
\begin{figure}[htb]
\centerline{
\epsfysize=10cm
\epsfbox{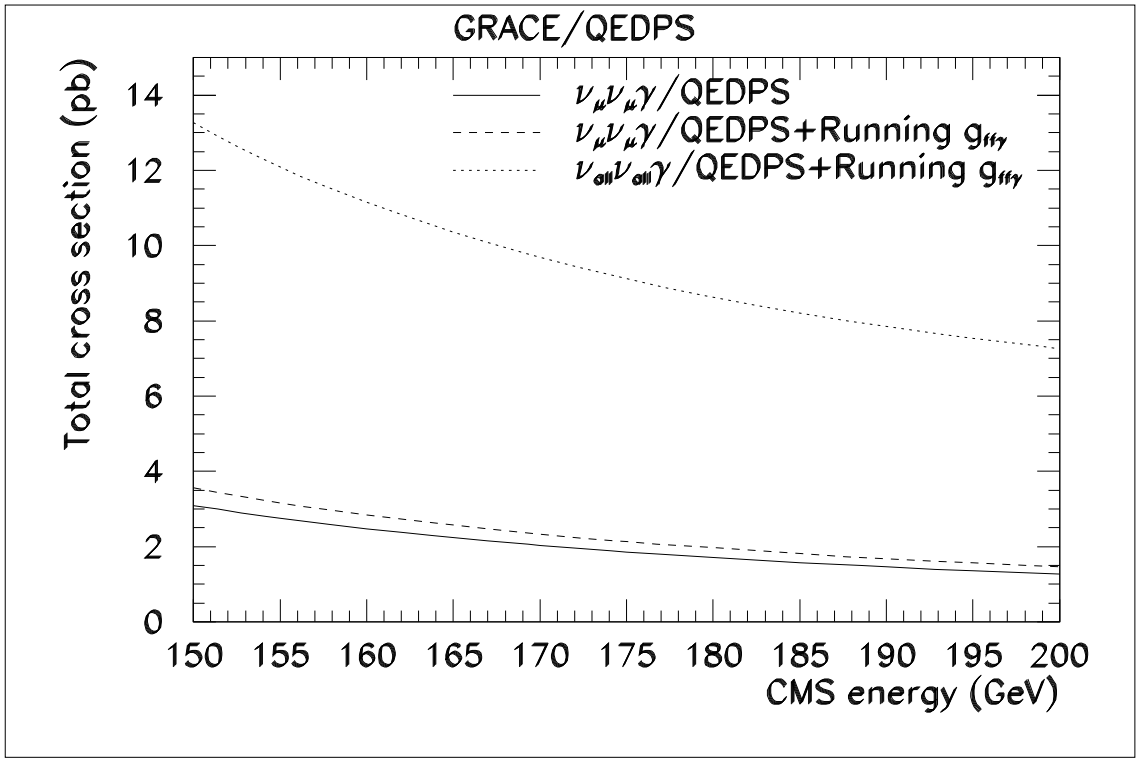}
}
\caption[FIG:roots2]{\label{FIG:roots2}\footnotesize
Total cross sections of the single-photon emission for the
$\nu_{\mu}$ and all neutrino processes 
using {\bf \tt grc$\nu\nu\gamma$} in $pb$.  
}
\end{figure}
\begin{figure}[htb]
\centerline{
\epsfysize=10cm
\epsfbox{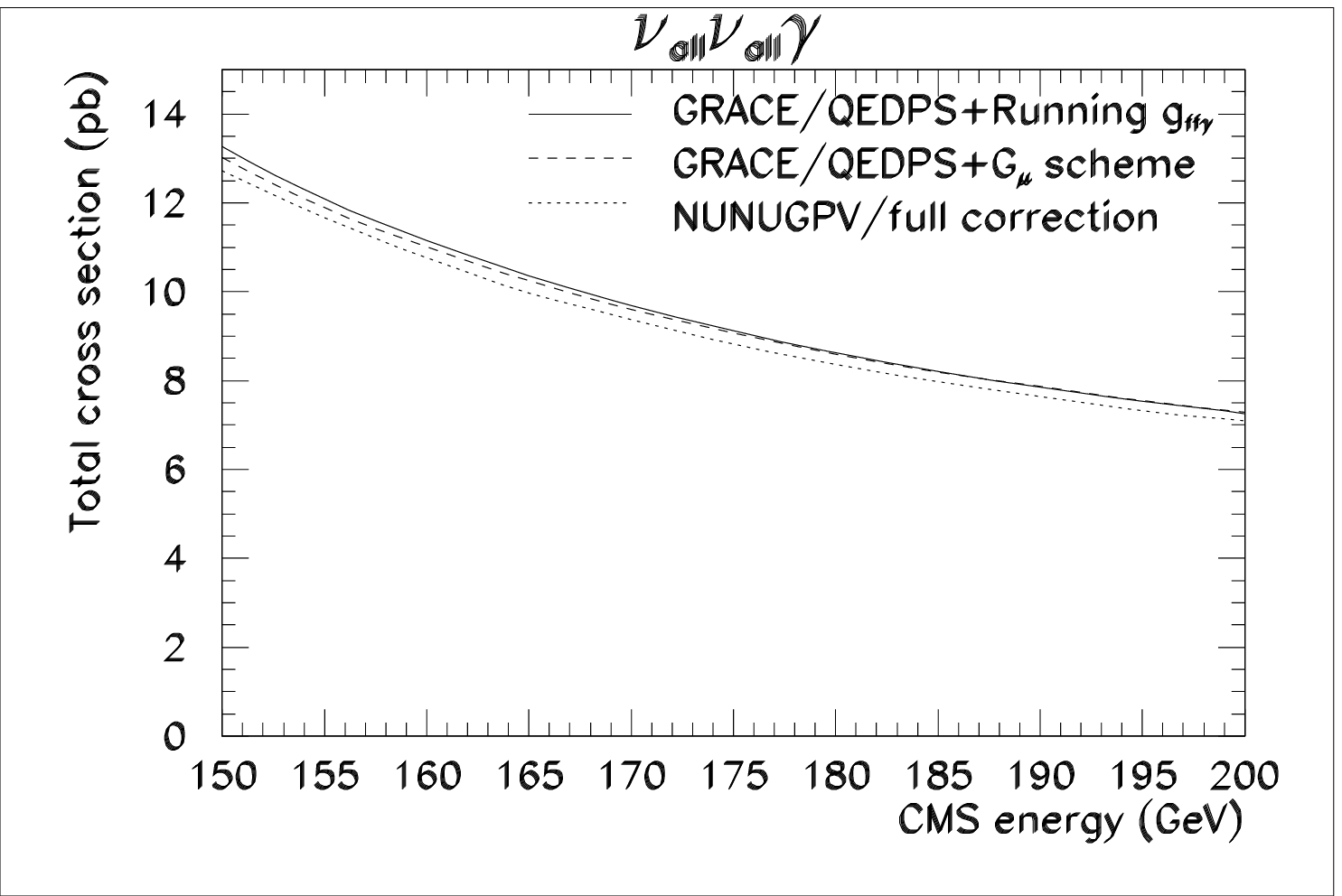}
}
\caption[FIG:roots3]{\label{FIG:roots3}\footnotesize
Total cross sections of the single-photon emission for 
all the neutrino processes using {\bf \tt grc$\nu\nu\gamma$} 
and {\tt NUNUGPV} in $pb$. The latter implements the $G_\mu$ scheme
for the higher order corrections. 
}
\end{figure}
\begin{figure}[htb]
\centerline{
\epsfysize=15cm
\epsfbox{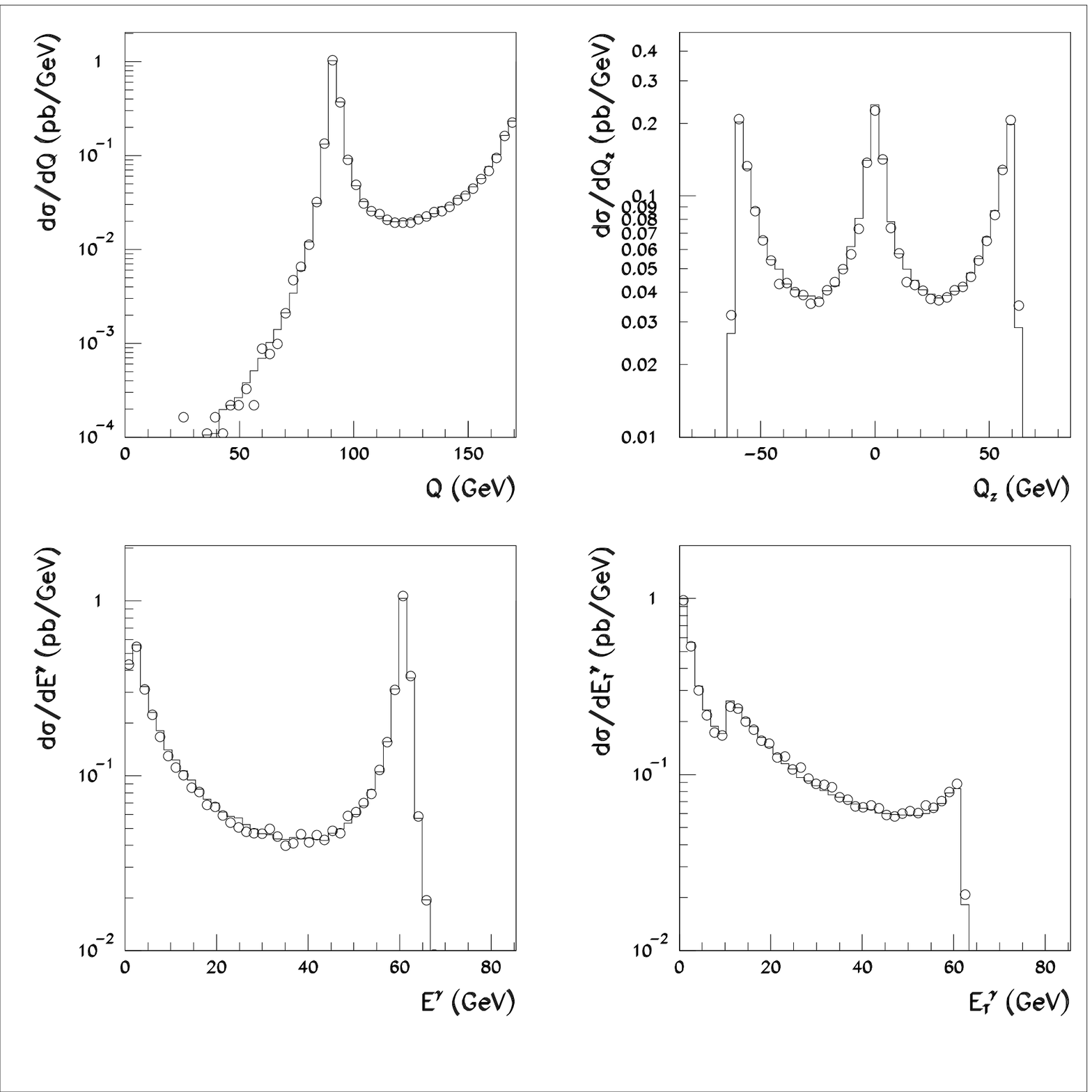}
}
\caption[FIG:spect3]{\label{FIG:spect3}\footnotesize
Differential cross sections with respect to the CMS energy of the
$\nu \bar{\nu}$ system,
its longitudinal component, energy and transverse energy of the hard photon
obtained by {\bf \tt grc$\nu\nu\gamma$} (solid histograms) 
and {\tt NUNUGPV} (circles) at the CMS
energy of 170 GeV.
}
\end{figure}
\begin{figure}[htb]
\centerline{
\epsfysize=10cm
\epsfbox{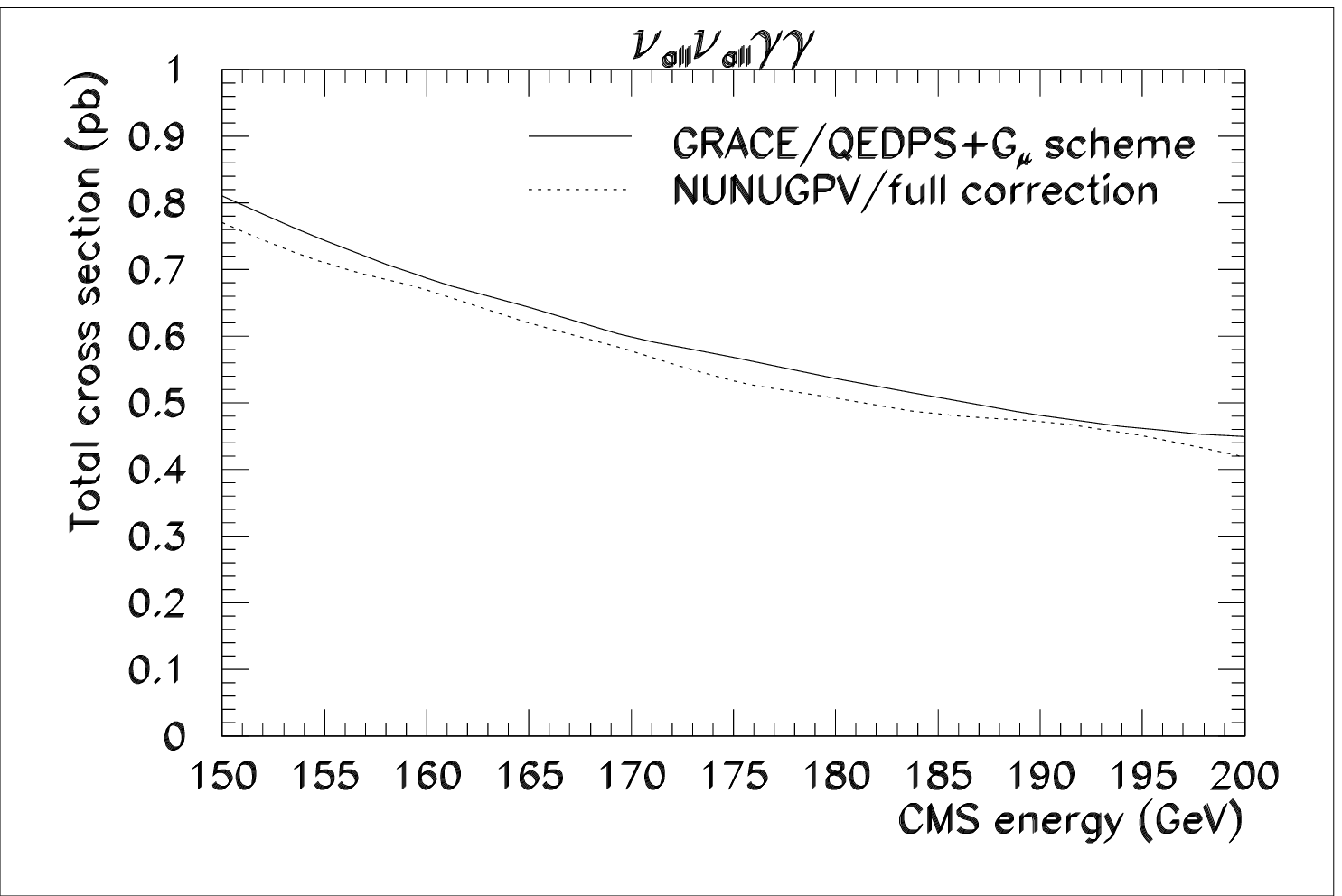}
}
\caption[FIG:roots4]{\label{FIG:roots4}\footnotesize
Total cross sections of the double-photon emission for 
all the neutrino processes using {\bf \tt grc$\nu\nu\gamma$} 
and {\tt NUNUGPV} in $pb$. The latter implements the $G_\mu$ scheme
for the higher order corrections. 
}
\end{figure}
\begin{figure}[htb]
\centerline{
\epsfysize=15cm
\epsfbox{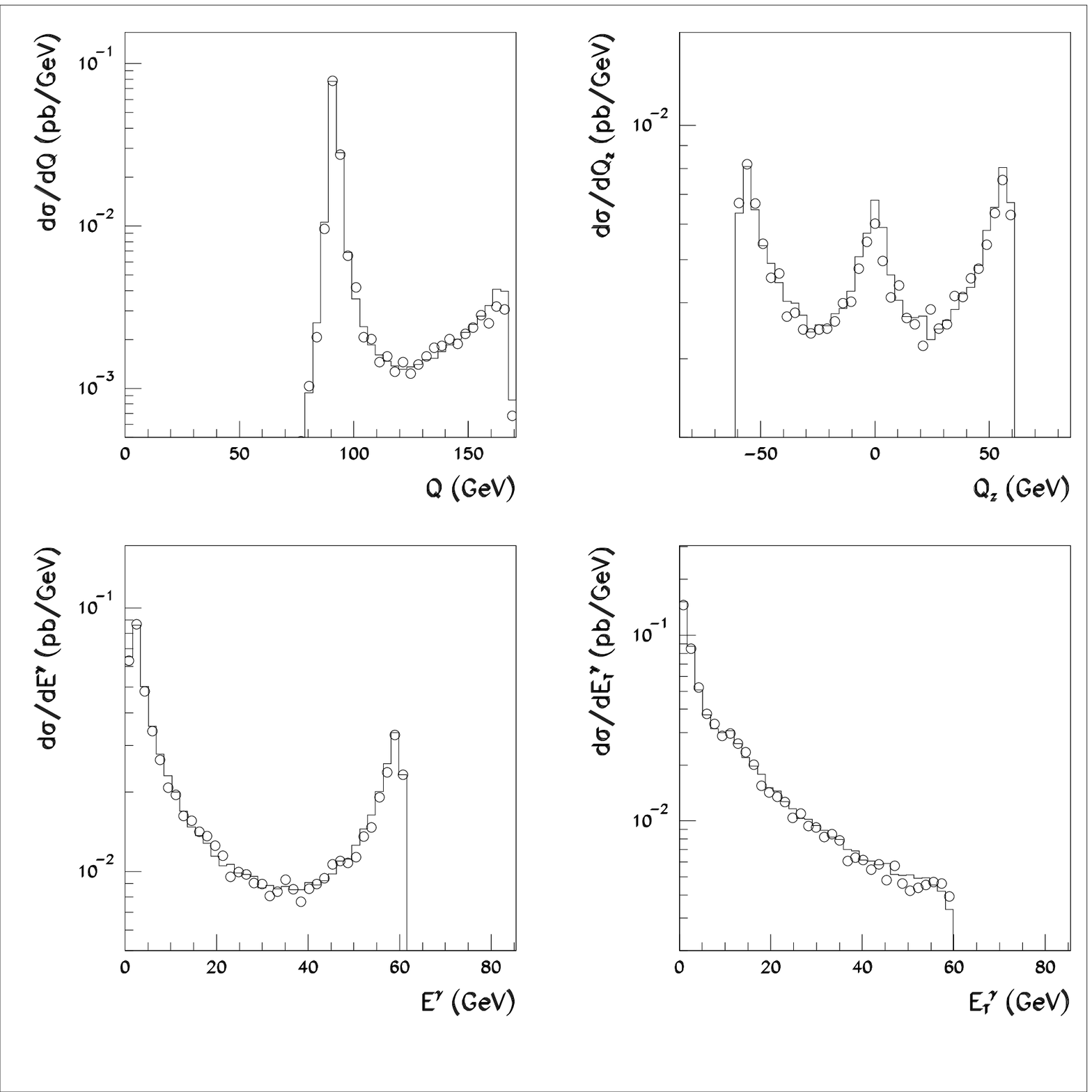}
}
\caption[FIG:spect4]{\label{FIG:spect4}\footnotesize
Differential cross sections with respect to the CMS energy of the
$\nu \bar{\nu}$ system,
its longitudinal component, energy and transverse energy of the hard photon
obtained by {\bf \tt grc$\nu\nu\gamma$} (solid histograms) 
and {\tt NUNUGPV} (circles) at the CMS
energy of 170 GeV.
}
\end{figure}
\begin{figure}[htb]
\centerline{
\epsfysize=15cm
\epsfbox{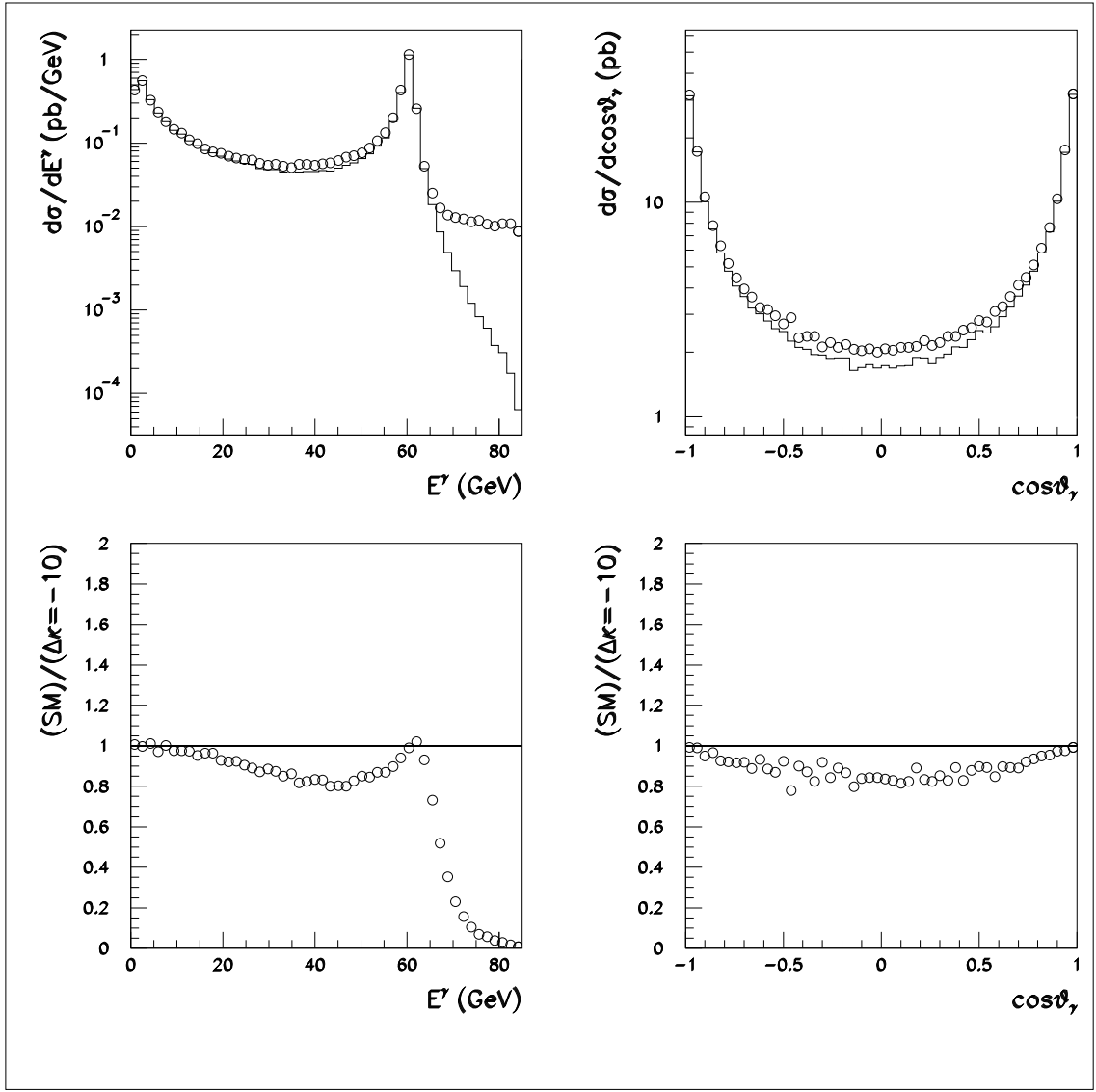}
}
\caption[FIG:atgc1]{\label{FIG:atgc1}\footnotesize
Energy and angular distributions of the hard photon
obtained by {\bf \tt grc$\nu\nu\gamma$} with the standard
TGC (solid histograms) 
and the anomalous TGC with $\Delta k_\gamma =-10$ at the CMS
energy of 170 GeV.
The distributions of the ratio between the standard and the anomalous TGC
are also shown.
}
\end{figure}
\begin{figure}[htb]
\centerline{
\epsfysize=10cm
\epsfbox{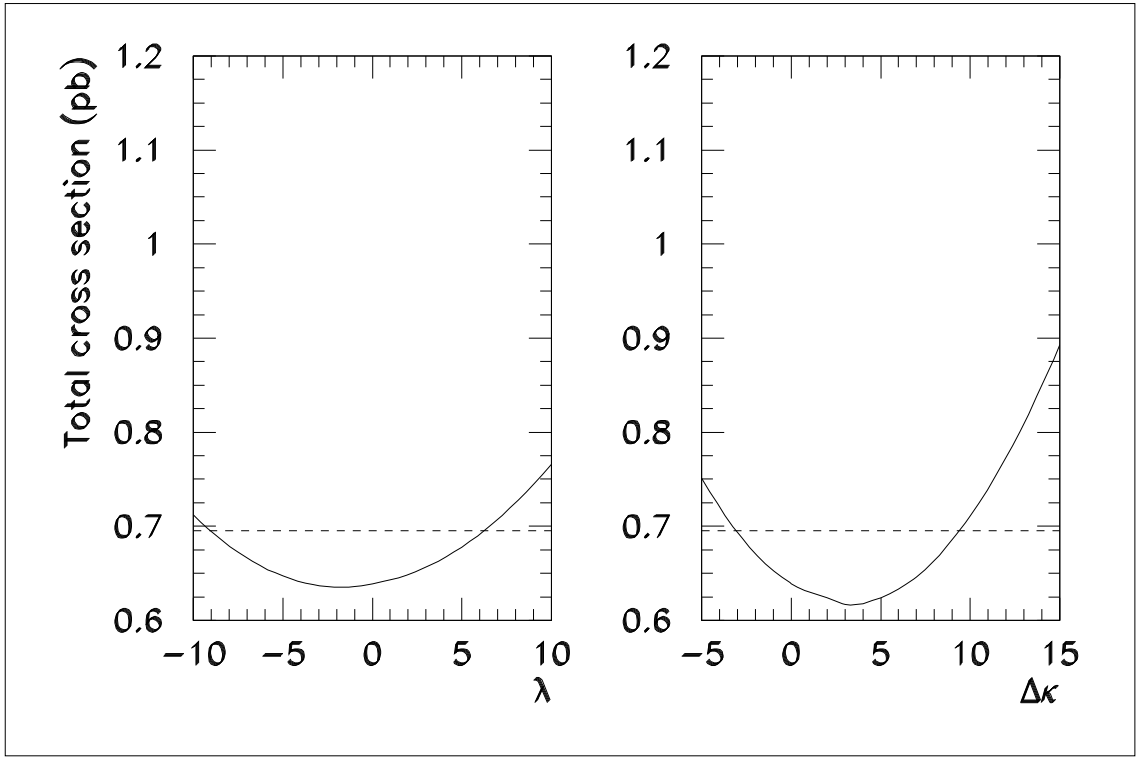}
}
\caption[FIG:atgc2]{\label{FIG:atgc2}\footnotesize
Total cross sections of the single hard-photon events as a function of
$\lambda_\gamma$ (left) and $\Delta \kappa_\gamma$ (right) at the CMS
energy of 170 GeV.
The dashed lines are the expected experimental upper limit at 
three standard-deviations of the statistical error 
from the standard model parameters after $200pb^{-1}$ data accumulation.
}
\end{figure}
\clearpage

\leftline{\Huge\bf Appendix}
\vskip 0.2cm
\leftline{\large\bf A.1 How to install the program}

The source code  is available by {\tt anonymous ftp} from
{\tt ftp.kek.jp} in the directory {\tt kek/minami/grcnna}.
The {\bf \tt grc$\nu\nu\gamma$} system contains the complete set 
of FORTRAN sources for 
$e^+e^- \rightarrow \nu \bar{\nu} \gamma$ and
$e^+e^- \rightarrow \nu \bar{\nu} \gamma \gamma$ matrix-elements and
the four libraries, i.e., 
{\tt BASES/SPRING}, {\tt CHANEL}, {\tt QEDPS} and
utilities for kinematics.
Those source codes are written in FORTRAN77.
{\bf \tt grc$\nu\nu\gamma$} has been developed on HP-UX, but should run on
any UNIX platform with a FORTRAN complier. 

The procedure of installation is as follows:
\begin{enumerate}
\item Editing {\it Makefile}.

The following macros in {\it Makefile} should be taken care of by 
users themselves.
For example, {\tt GRCNNADIR} defines the directory name
where {\bf \tt grc$\nu\nu\gamma$} is installed.
The values of {\tt FC} and {\tt FOPT} define the relevant compiler
name and option for your system.
The other macros can be left as they are.

\begin{tabular}{lcl}
  {\tt GRCNNADIR} &{\tt =}& directory where {\bf \tt grc$\nu\nu\gamma$} 
are installed. \\
  {\tt LIBDIR}   &{\tt =}& directory where libraries are installed. \\
 & &  (default is \verb+$(GRCNNADIR)/lib+.) \\
  {\tt BINDIR}   &{\tt =}& directory where an executable is installed. \\
 & & (default is \verb+$(GRCNNADIR)/bin+.) \\
  {\tt MACHINE} &{\tt =}& {\tt [hpux|hiux|sgi|dec|sun]} \\
  {\tt FC}      &{\tt =}& FORTRAN compiler command name. \\
  {\tt FOPT}    &{\tt =}& FORTRAN compiler options. \\
  {\tt CERNLIBS} &{\tt =}& CERNLIB including the jetset library. \\
& & For HP-UX, \verb+-L/cern/pro/lib -ljetset74+ \\
& & \verb+-lpacklib -lkernlib -L/lib/pa1.1/ -lm+ \\
\end{tabular}

\item Compilation.

  By executing command {\tt make install} the executable of the interface
  program is generated at {\tt BINDIR}.
  Furthermore four libraries, {\tt BASES/SPRING}, {\tt CHANEL}, {\tt QEDPS}
  and kinematics utility library,
  are generated in {\tt LIBDIR}.

%
\end{enumerate}
\newpage

\leftline{\large\bf A.2 How to run the program}

According to the number of hard-photons, two directories
are prepared: {\bf \tt \$(GRCNNADIR)/1a} and {\bf \tt \$(GRCNNADIR)/2a}.
Three FORTRAN files and a {\tt Makefile} are stored in each directory.

For the numerical integration, {\tt BASES}:
\begin{itemize}
\item {\bf \tt mainbs.f}: A main program of the numerical integration.
\item {\bf \tt usrprm.f}: A subroutine for parameter setting.
\end{itemize}
For the event generation, {\tt SPRING}:
\begin{itemize}
\item {\bf \tt mainsp.f}: A main program of the event generation. Event-loop
is also included. 
\end{itemize}
The subroutine, {\bf \tt usrprm.f}, is used commonly for both of {\tt BASES}
and {\tt SPRING}.

Users can set the parameters of the calculations by editing the file
{\bf \tt usrprm.f}. The key words are as follows;

\par\smallskip
\begin{verbatim}

* CM energy
      E0   = 170.d0
* Hard Photon condition
      Ea   =  1.d0
      THEa = 10.d0

* channel selection
*     Ineut=1   : sum up all neutrino species
*          =2   ; nu-e only
*          =3   : nu-mu (nu-tau )only
      Ineut = 1

* QED correction *********************************************
* QEDPS switch                                               !
*     Ips=0     : No ISR                                     !
*        =1     : ISR by QEDPS                               !
      Ips=1                                                  !
* QED correction *********************************************

* Higher order correction scheme *****************************
* Running alpha                                              !
*     Iruna=0   : On-shell scheme ; fixed alpha              !
*          =1   : On-shell scheme ; running alpha            !
*          =3   : Gmu scheme                                 !
      Iruna=1                                                !
*                                                            !
* Running width                                              !
*     Irunw=0   : fixed width                                !
*          =1   : running width                              !
      Irunw=1                                                !
                                                             !
* Set QED alpha                                              !
      alpha0= 1.0d0/137.0359895d0                            !
      if(Iruna.ne.1) then                                    !
        alpha = 1.0d0/128.07d0                               !
       else                                                  !
        alpha = 1.0d0/137.0359895d0                          !
       end if                                                !
* Higher order correction scheme *****************************

* WWgamma coupling
*
*     ang1a :     g1   =1 (SM)
*     andka : delta-k  =0 (SM)
*     anlma : lambda   =0 (SM)
      ang1a    =1
      andka    =0
      anlma    =0

\end{verbatim}
\par\smallskip\noindent
The parameters for the anomalous TGC are available only for the single-photon
case. 

Users can proceed with the calculations as follows:
\par
\begin{itemize}
\item[i)] Change directory by typing
\begin{verbatim}
% cd $(GRCNNADIR)/1a
\end{verbatim}
or
\begin{verbatim}
% cd $(GRCNNADIR)/2a
\end{verbatim}
\item[ii)] Edit {\bf \tt usrprm.f} to set user parameters.
\item[iii)] Create an executable {\tt integ} for the integration by typing
\begin{verbatim}
% make integ
\end{verbatim}
\par
\item[iv)] Numerical integration is actually performed by typing
\begin{verbatim}
%  integ
\end{verbatim}
\par
The results of
the integration step are displayed on the console as well as written in an
output file 
{\tt bases.result}. The total cross section in $pb$ and the
estimated statistical error are shown 
on the last line, under {\tt Cumulative Result}, 
in the table of the {\tt Convergence Behavior for the Integration step}.
The differential cross sections are also printed as a function of the energy, 
scattering angle of each particle and invariant masses of any two final particles. 
The probability distribution of the integrand is written in a file 
{\tt bases.data} which will be used in the event generation step by {\tt spring}.

\item[v)] Before running the event generation, users may edit 
{\tt mainsp.f} to set additional parameters if needed 
and call the user's own analysis routines.

The following is the structure of the generated {\tt mainsp.f}, where
four-momentum of all particles including soft-photons generated by QESPS
are stored in the {\tt common/lujets/} in the 
{\tt JETSET} format when subprogram {\tt sp2lnd} is called in the event-loop:
\begin{verbatim}
      Program  mainsp
      implicit real*8(a-h,o-z)
      external func
       ....................
      real*4  p,v
      common/lujets/n,k(4000,5),p(4000,5),v(4000,5)
       ....................
       ....................
      mxtry  =   50
      mxevnt = 1000
      do 100 nevnt = 1, mxevnt

         call spring( func, mxtry )
       ....................
*         -----------------
           call sp2lnd
*         -----------------
*
*       ==============================================
*       (  user_analysis based on the common lujets )
*       ==============================================
*
  100 continue
       ....................
      stop
      end
\end{verbatim}

\item[iv)] Create an executable {\tt spring} for event generation
by typing

\begin{verbatim}
% make spring
\end{verbatim}

\item[vii)] Start the event generation by typing

\begin{verbatim}
% spring
\end{verbatim}

Information concerning the event generation will be written in the
{\tt spring.result} file. Users should pay attention to the 
histograms generated in this step. 
The distributions of the generated events are superimposed
with the character ``{\tt 0}'' on the histograms generated in
the integration step. These two
distributions should be consistent with each other 
within the statistical error of
the generation. For the details of the output files of {\tt BASES}
and {\tt SPRING},
users can consult Ref.\cite{BASES}. 
\end{itemize}
\end{document}